\documentclass{article}
\usepackage[preprint, nonatbib]{neurips_2020}

\usepackage[utf8]{inputenc} % allow utf-8 input
\usepackage[T1]{fontenc}    % use 8-bit T1 fonts
\usepackage{hyperref}       % hyperlinks
\usepackage{url}            % simple URL typesetting
\usepackage{booktabs}       % professional-quality tables
\usepackage{amsfonts, amsmath, bm}       % blackboard math symbols
\usepackage{nicefrac}       % compact symbols for 1/2, etc.
\usepackage{microtype}      % microtypography
\usepackage{amsmath,amssymb}
\usepackage{graphicx}

\title{Unsupervised identification of rat behavioral motifs across timescales}

\author{
  Haozhe Shan \\
  Harvard University\\
  Cambridge, MA, 02138 \\
  \texttt{hshan@g.harvard.edu}
  \And
   Peggy Mason \\
   University of Chicago \\
   Chicago, IL, 60615 \\
   \texttt{pmason@uchicago.edu}
}

\begin{document}

\maketitle

\begin{abstract}
Behaviors of several laboratory animals can be modeled as sequences of stereotyped behaviors, or behavioral motifs. However, identifying such motifs is a challenging problem. Behaviors have a multi-scale structure: the animal can be simultaneously performing a small-scale motif and a large-scale one (e.g. \textit{chewing} and \textit{feeding}). Motifs are compositional: a large-scale motif is a chain of smaller-scale ones, folded in (some behavioral) space in a specific manner. We demonstrate an approach which captures these structures, using rat locomotor data as an example. From the same dataset, we used a preprocessing procedure to create different versions, each describing motifs of a different scale. We then trained several Hidden Markov Models (HMMs) in parallel, one for each dataset version. This approach essentially forced each HMM to learn motifs on a different scale, allowing us to capture behavioral structures lost in previous approaches. By comparing HMMs with models representing different null hypotheses, we found that rat locomotion was composed of distinct motifs from second scale to minute scale. We found that transitions between motifs were modulated by rats' location in the environment, leading to non-Markovian transitions. To test the ethological relevance of motifs we discovered, we compared their usage between rats with differences in a high-level trait, prosociality. We found that these rats had distinct motif repertoires, suggesting that motif usage statistics can be used to infer internal states of rats. Our method is therefore an efficient way to discover multi-scale, compositional structures in animal behaviors. It may also be applied as a sensitive assay for internal states.

\end{abstract}

\section*{Introduction}

To survive, animals must adapt to and interact with constantly changing environments. On the other hand, their behavioral repertoires  are limited by factors such as biomechanics and neural circuit complexity. As behavioral scientists have observed, animals often exhibit stereotyped behaviors, or behavioral motifs \cite{dantzer1986behavioral}. Dynamic behavioral adaptation may thus arise from assembling motifs in different ways to serve different ends. 

\begin{figure}
    \includegraphics[width=1\textwidth]{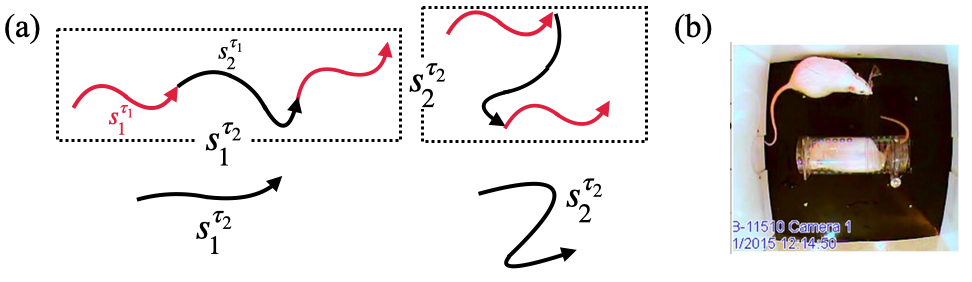}
    \caption{(a) The "folding problem" in motif identification. The same sequence of motifs at timescale $\tau_1$, $(s^{\tau_1}_1, s^{\tau_1}_2, s^{\tau_1}_1)$, may have different spatial structures (top), corresponding to two different motifs, $s^{\tau_2}_1, s^{\tau_2}_2$ at timescale $\tau_2>\tau_1$; the two $\tau-2$ motifs are indistinguishable one simply describes them as sequences of $\tau_1$ motifs. Our approach removes smaller-scale information and preserves the spatial structures (bottom). (b) Setup of behavioral experiments where data were collected. The rat in the center was trapped in the restrainer and could not move; Locomotion of the rat outside was tracked and used for all analyses.} 
    \label{fig:motifs}
\end{figure}

Identifying motifs in animal behaviors presents a difficult machine learning problem. Here, we discuss four main challenges. (1) Behavioral motifs are not demarcated by clear indicators. An analogy can be made with the tokenization problem in East Asian languages. There, the lack of spaces between characters makes it difficult to segment a stream of characters into meaningful words. Similarly, it is difficult to segment a stream of coordinates in some behavioral space into meaningful motifs. (2) Motifs have a multi-scale, compositional structure. Motifs may simultaneously exist at different spatial and temporal scales: an animal amid a \textit{chewing} motif may also be in a \textit{feeding} motif, depending on the spatial and temporal scale one is concerned with. In fact, small-scale motifs may compose large-scale motifs. (3) Some large-scale motifs are only distinguished by their structures, which are not captured by the temporal ordering of small-scale motifs composing them. As an analogy, amino acids (small-scale motifs) are linked to form proteins (large-scale motifs); yet proteins are not fully described by their amino-acid chains, because the same chain can undergo different folding. We illustrate this problem in Fig.\ref{fig:motifs}(a). Here, $\tau_2>\tau_1$ are two timescales. Directed curves are motifs in some behavioral space. Both large scale motifs $s_1^{\tau_2}, s_2^{\tau_2}$ are composed of the same sequence of small-scale motifs, yet arranged into different spatial structures Fig.\ref{fig:motifs}(a).  We call this the "folding problem" in motif identification. (4) Labelled data is extremely limited. While humans can score a few well-studied motifs (e.g. feeding and grooming in rodents \cite{van2001validation}), these are likely a small fraction of the animal's full behavioral repertoire \cite{wiltschko2015mapping}. Such human-labeled data are therefore of little use for mapping the entire behavioral repertoire in animals. One is thus restricted to unsupervised learning methods. This also makes it difficult to evaluate models, since one cannot simply compute the accuracy using labeled data. A potential solution to this problem, which we used in the present work, is to see whether identified motifs are related to some ethologically meaningful construct, such as the animal's trait or internal state. 

Recently, there have been several studies on the motif identification problem \cite{wiltschko2015mapping,berman2014mapping,brown2013dictionary, tao2019statistical}. These studies applied unsupervised learning methods (often directed acyclic graphical models) to overcome the first challenge (demarcating motifs) we identified above. They were able to discover motifs in animal behaviors previously unknown to human observers. However, the majority of these studies focused on motifs of sub-second timescales \cite{wiltschko2015mapping,berman2014mapping,brown2013dictionary}. This approach has two limitations. First, sub-second timescales are too short to capture many, if not most, behaviors of interest in some laboratory animals. In common behavioral paradigms for rodents, such as the elevated plus maze (EPM), open field test, forced swim test and so on \cite{detke1995active,Hogg1996}, supra-second scale navigation actions are relevant. Beyond these well established paradigms, other stereotyped behaviors in rodents that have been identified are also of larger scale. Some examples are boxing and lateral threat for social aggression \cite{blanchard1975conspecific}, thigmotaxis for anxiety \cite{treit1988thigmotaxis}, and licking and grooming for maternal care \cite{pedersen1985oxytocin}. Second, since these studies focused on a single timescale, they did not capture multiscale, compositional structures in behaviors. To overcome this limitation, Tao and colleagues\cite{tao2019statistical} applied Hierarchical Hidden Markov Models (HHMM) to motif identification. As the name suggests, in HHMMs, motifs are arranged in a hierarchy. As a conceptual example, both \textit{sprinting} and \textit{lunging} may fall under the higher-level motif, \textit{hunting}. Since there are fewer high-level motifs, they transition less frequently and thus capture structures on a larger timescale. One issue with this approach is that the model assumes exclusivity of motifs. That is, each small-scale motif has to fall under one or another large-scale motif, but not both. Clearly, motifs such as \textit{running} can be general-purposed, serving multiple large-scale motifs. In addition, HHMMs do not resolve the folding problem. Since they treat large-scale motifs as sequences of smaller-scale motifs, the folding information is lost.

In the current work, we demonstrate a method to overcome all four challenges discussed above for motif identification. As an example, we used rat locomotor data (see setup in Fig.\ref{fig:motifs}(b)), where the behavioral space is naturally the physical space. Our approach was to train multiple unsupervised models (Hidden Markov Models, HMM) in parallel, forcing each to learn motifs on a different scale. We first binned the raw locomotor trajectories into segments at different levels of granularity (bin sizes range from 1 to 120 seconds). For segments at each level, we performed dimensionality reduction, which naturally eliminated finer-scale structures in the behavior. We then trained a separate HMM for segments on each level. For each HMM, the data it "saw" only contained motifs on the scale specified by the bin size. This essentially forced each HMM to capture motifs on a different scale. Compositionality naturally arose as each larger-scale motif overlapped in time with several smaller-scale motifs. Since the large-scale segments were not simply a sequence small-scale ones, they preserved the "folding" structures (Fig.\ref{fig:motifs}(a)). To properly evaluate whether the motifs our models discovered were ethological meaningful, we compared motif usage between rats with known differences in prosociality. Difference in this trait had been confirmed experimentally using a paradigm for rat prosocial behaviors\cite{bartal2011empathy,bartal2014pro,bartal2016anxiolytic,shan2016rodent, carvalheiro2019helping, tomek2019effects}. Both motif usage probabilities and transition probabilities were different between prosocial and non-prosocial rats, further validating the ethological relevance of motifs we discovered. 

Our model is the first to capture behavioral motifs in lab animals on timescales from second-level to minute-level to the best of our knowledge. The parallel HMM approach we used allowed easy parameter estimation and captures compositionality and spatial structures. While we focused on locomotor behaviors in rats, our approach is readily extended to behavioral data from other laboratory animals.

\section*{Materials and methods}
Codes used for all analyses are available on Github (github.com/hzshan/HMM). Behavioral experiments that provided data for the current analysis were conducted in accordance with established ethical standards and were approved by the University of Chicago Institutional Animal Care and Use Committee.

%%%%%%%%%%%%%%%%%%%%%%%%%%%%%%%%%%%%%%%%%%%%%%%%%%%%%%%
%%%%%%%%%%%%%%%%%%%%%%%% FIGURE 2 %%%%%%%%%%%%%%%%%%%%%
%%%%%%%%%%%%%%%%%%%%%%%%%%%%%%%%%%%%%%%%%%%%%%%%%%%%%%%
\begin{figure}
    \includegraphics[width=1\textwidth]{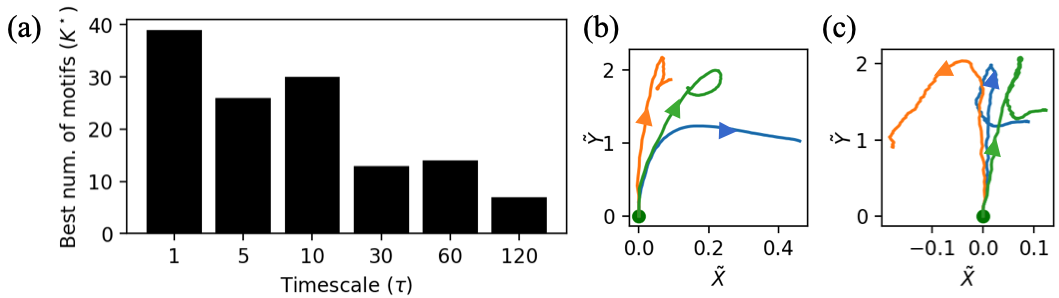}
    \caption{(a) For each timescale $\tau$, the best-fitting number of motifs ($K^\star$) was selected using BIC. In general, the number of motifs was smaller for larger timescales. (b) Three example motifs, shown in egocentric coordinates (in centimeters), for $\tau=5$; (c) Same as (b), for $\tau=10$.}
    \label{fig:model_comparison}
\end{figure}

\begin{figure}
    \includegraphics[width=0.8
    \textwidth]{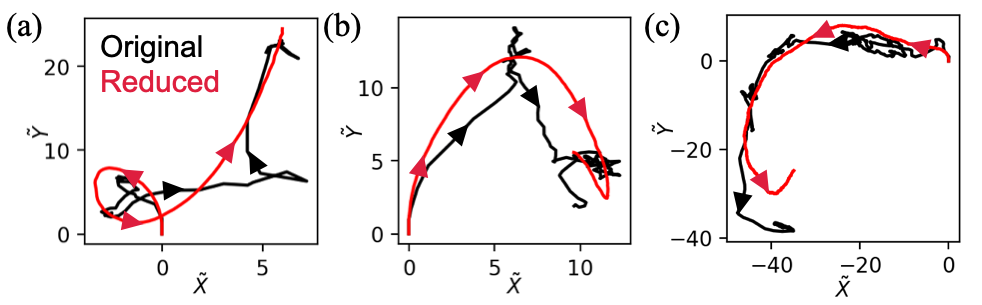}
    \caption{Dimensionality reduction removed smaller-scale structures, allowing each HMM to learn motifs of a different scale. (a) An example of an original $\tau=10$ trajectory (black) and its dimensionality-reduced counterpart (i.e. reconstructed from PCA, red). Coordinates are in centimeters. (b)(c) Same as (a), but for $\tau=30, 60$.}
    \label{fig:dim reduction}
\end{figure}

\subsection*{Experimental setup}
The basic paradigm for the behavioral experiments has been described in detail before \cite{bartal2011empathy,bartal2014pro,bartal2016anxiolytic,shan2016rodent,carvalheiro2019helping, tomek2019effects}. In short, adult rats were allowed to roam freely in arenas containing a centrally located trapped rat in sessions of 60 minutes. Locomotor behaviors were recorded with overhead CCD cameras and tracked by the software Ethovision \cite{noldus2001ethovision}. 24 free rats were tested for 12 sessions, each with 35 minutes of analyzed tracking at 15 frames per second, resulting in ~9 million tuples of coordinates.

\subsection*{Data description, preprocessing, and model setup}

The raw data contained the 2D allocentric coordinates of each rat over time. We binned the data with bin sizes $\tau$ of 1, 5, 10, 30, 60, and 120 seconds, creating six versions of the same dataset. We did not use larger bins because that would reduce the number of datapoints (i.e. bins) too much for parameter estimation. The bin size $\tau$ effectively dictated the timescale of motifs each dataset described. For data binned with each size, we treated the sequence of coordinates in each bin as one observation for the Hidden Markov Model. 

Formally, let $\{\bm{x}_t\}_{t=1...T}$ be the raw data from each recording session, where $\bm{x}_t=(X,Y)$ are the 2D allocentric coordinates for the tracked rat. Since recording was performed at 15 frames per second, $T/15$ is the duration of this session in seconds. We segmented the data into spans of length $\tau$, and obtained $\{(\bm{x}_1, \bm{x}_2,...,\bm{x}_\tau), (\bm{x}_{\tau+1}, \bm{x}_{\tau+2},...,\bm{x}_{2\tau}),... \}$. Each span, $o^\tau_i=(\bm{x}_{i\tau}, \bm{x}_2,...,\bm{x}_{(i+1)\tau})$, is hereafter referred to as the $i$-th \textsc{trajectory}. It constitutes the \textsc{observation} in HMM literature. 

Before estimating parameters for HMMs, we performed two preprocessing procedures on the trajectories. We shifted the frame of reference to an egocentric one; The first coordinate of each trajectory was shifted to $(0,0)$; the trajectory was then rotated such that the second coordinate lay on the positive $y$-axis, $(0,y)$. These procedures collapsed translational and rotational symmetries in the trajectories. After alignment, we treated each trajectory as a $2\times\tau$ dimensional vector and performed dimensionality reduction with principle component analysis (PCA). Reduced vectors are 5-dimensional and capture more than 90\% of total variance. We denote the processed trajectory $\tilde{o^\tau_i}$. For trajectories at each scale, this procedure removed smaller-scale structures and preserved the structure of interest (Fig.\ref{fig:dim reduction}).

In HMMs, each $\tilde{o^\tau_i}$ is assumed to be the noisy version of a \textsc{motif} $s^{\tau}_i$, and there are a total of $K^{\tau}$ motifs on the timescale $\tau$. Motif transitions are assumed to be Markovian. These assumptions can be summarized as

\begin{align}
    & p(\tilde{o^\tau_i}|s^{\tau}_1,s^{\tau}_2,...,s^{\tau}_{i},\tilde{o^\tau_1},...,\tilde{o^\tau_{i-1}})=p(\tilde{o^\tau_i}|s^{\tau}_{i}) \\
    & p(s^{\tau}_{i}|s^{\tau}_1,s^{\tau}_2,...,s^{\tau}_{i-1},\tilde{o^\tau_1},...,\tilde{o^\tau_{i-1}})=p(s^{\tau}_{i}|s^{\tau}_{i-1}).
\end{align}

The conditional probabilities $p(\tilde{o^\tau_i}|s^{\tau}_{i})$ and $p(s^{\tau}_{i}|s^{\tau}_{i-1})$ are parameters to be estimated from data. 

To determine whether distinct motifs and transitions existed in the data, we compared the performance of HMMs to two other models. The first, a Gaussian Model (GM), assumes a single motif for each timescale, i.e. 

\begin{equation}
    p(\tilde{o^\tau_i}|s^{\tau}_1,s^{\tau}_2,...,s^{\tau}_{i},\tilde{o^\tau_1},...,\tilde{o^\tau_{i-1}})=p(\tilde{o^\tau_i});
\end{equation}
The second model is a Gaussian Mixture Model (GMM), where multiple motifs are assumed to exist, but there is no motif transition structure, i.e. 
\begin{align}
    & p(\tilde{o^\tau_i}|s^{\tau}_1,s^{\tau}_2,...,s^{\tau}_{i},\tilde{o^\tau_1},...,\tilde{o^\tau_{i-1}})=p(\tilde{o^\tau_i}|s^{\tau}_{i}) \\
    & p(s^{\tau}_{i}|s^{\tau}_1,s^{\tau}_2,...,s^{\tau}_{i-1},\tilde{o^\tau_1},...,\tilde{o^\tau_{i-1}})=p(s^{\tau}_{i}).
\end{align}

Model comparisons and hyperparameter selection were performed on held-out data (30\% of the dataset) using Bayesian Information Criterion (BIC) as a metric.

%%%%%%%%%%%%%%%%%%%%%%%%%%%%%%%%%%%%%%%%%%%%%%%%%%%%%%%
%%%%%%%%%%%%%%%%%%%%%%%% RESULTS  %%%%%%%%%%%%%%%%%%%%%
%%%%%%%%%%%%%%%%%%%%%%%%%%%%%%%%%%%%%%%%%%%%%%%%%%%%%%%
\section*{Results}
At the timescales considered in the current analysis (1 to 120 seconds), HMMs and GMMs were both favored over GMs. This rejected the null hypothesis that on each scale, all trajectories arose from the same motif. In other words, the trajectories formed distinct clusters. For HMMs and GMMs, a hyperparameter to be tuned was the number of motifs $K$. Again using BIC on held-out data as a metric, we selected the best $K$, $K^\star$, for HMMs and GMMs of each timescale. We found that across the$\tau$s considered, HMMs consistently outperformed GMMs, suggesting non-trivial transition structures between motifs. We illustrate the best-fitting number of motifs $K^\star$ in Fig.\ref{fig:model_comparison}(a). To determine how many behavioral motifs existed at each timescale, multiple HMMs and GMMs with different values for $K$ were fitted over data and $K$ with the highest BIC was selected. Some example motifs are shown in Fig.\ref{fig:model_comparison}(b)(c).

Our model comparison produced evidence that distinct motifs and Markovian transition between them existed at an up-to-a-minute timescale in rat locomotion. We proceeded to analyze properties of motifs and their transitions.

%%%%%%%%%%%%%%%%%%%%%%%%%%%%%%%%%%%%%%%%%%%%%%%%%%%%%%%
%%%%%%%%%%%%%%%%%%%%%%%% FIGURE 3 %%%%%%%%%%%%%%%%%%%%%
%%%%%%%%%%%%%%%%%%%%%%%%%%%%%%%%%%%%%%%%%%%%%%%%%%%%%%%

\begin{figure}
\includegraphics[width=0.6\textwidth]{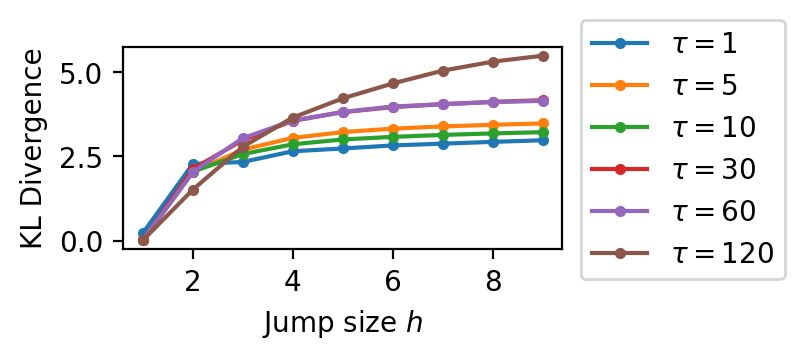}
\caption{The KL divergence of the observed $p(s^\tau_{i+h})$ from that predicted using $p(s^\tau_{i})$ and the Chapman-Kolmogorov equality. These results suggest that motif transition is not exactly Markovian; it becomes less Markovian as timescale increases.}
\label{fig:markov property}
\end{figure}

\subsection*{Motif transitions are not perfectly Markovian}
The HMMs we used assume Markovian transitions between motifs. While consistent with previous studies, this is likely simplistic. Animals may use their entire behavioral history to select the next behavior to perform. We therefore tested how well Markov processes, as assumed by HMMs, approximated motif transitions in rats.

One signature of Markovian processes is that the probability distribution of motifs at time $i+h$, $p(s^\tau_{i+h}$, can be computed exactly using $p(s^\tau_{i})$ and the Chapman-Kolmogorov equality. If motif transitions were not Markovian, then the observed $p(s^\tau_{i+h}$ would be different from the distribution computed in this way. We quantified this difference using the Kullback-Leibler (KL) divergence. We computed a deviation score $D(h)$ defined as
\begin{align}
    D(h)&=\sum_{s^\tau_{i+h}} p \left(s^\tau_{i+h}\right) \log \frac{p \left(s^\tau_{i+h}\right)}{\tilde{p} \left(s^\tau_{i+h}\right)} \\
    \tilde{p} \left(s^\tau_{i+h}\right) &= T^h p \left(s^\tau_{i}\right),
\end{align}
where $p \left(s^\tau_{i}\right)$ is a vector encoding marginal probabilities of motifs at time $i$, and $T^h$ is the transition probability matrix to the $h$-th power. A higher divergence indicates stronger deviation from Markovian transitions. The results for all $\tau$ are shown in Fig.\ref{fig:markov property}. As expected, the deviation increases and then plateaus as $h$ increases. Interestingly, deviation from Markovian transitions was stronger for longer timescales. This has an intuitive explanation -- larger-timescale motifs may reflect goal-oriented behaviors where rats make decisions based on their behavioral history.

\subsection*{Quantifying the effect of environment on behavior}

One potential approach to understanding non-Markovian motif transitions is to identify non-behavioral variables on which transitions depend. For instance, the motif \textit{running} may lead to a long-term increase of the hidden variable \textit{fatigue}, which in turn depresses the transition probabilities from other motifs to \textit{running}. However, we did not perform non-behavioral measurements. Fortunately, we were able to find one variable that might affect motif transition -- the animal's location in the arena.

\begin{figure}
\centering
\includegraphics[width=1\linewidth]{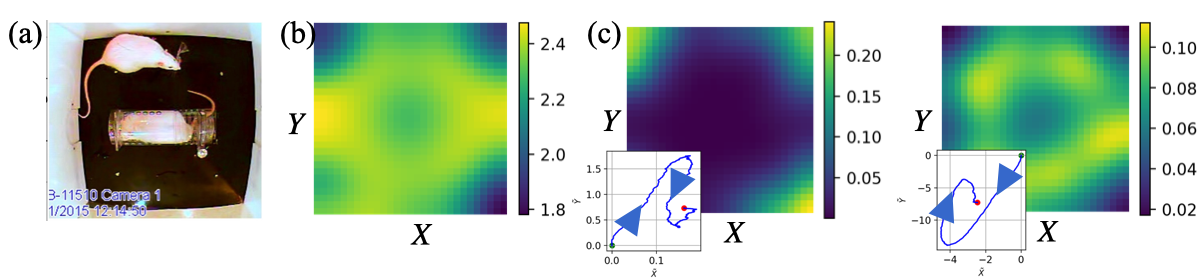}
\caption{Motif usage had environmental dependence. (a) Same as Fig.\ref{fig:motifs}(b) for reference. (b) Entropy of $p(s^\tau|\vec{x})$ for $\tau=30$. $\vec{x}=(X,Y)$ are the arena-centric coordinates (in centimeters). (c) Two example motifs for $\tau=30$ and their $p(s^\tau|\vec{x})$. Heatmaps show $p(s^\tau|\vec{x})$ for each motif over space; insets show the trajectories of these motifs in rat-centric coordinates ($(\tilde{X},\tilde{Y}$)).}
\label{fig:spatial heterogeneity}
\end{figure}

Since the trajectories fitted by HMMs were in egocentric coordinates, they did not contain information about where the rats were in the arena. On the other hand, there were good reasons to believe that our environment, while not naturalistic, was ethologically salient and might strongly affect motif transitions (Fig.\ref{fig:spatial heterogeneity} (a)). The trapped conspecific in the center of the arena is known to elicit stress responses in the tracked rat \cite{bartal2011empathy}. On the other hand, rats are known to prefer the safety of edges of the arena -- this preference, known as thigmotaxis, is in fact used to measure anxiety-like states in rats \cite{treit1988thigmotaxis}. Therefore, if motif usage displayed location specificity, it reflected different ethological meanings for different motifs.

First, we asked whether diversity of used motifs depended on location in the arena. We measured behavioral diversity at each allocentric location $\bm{x}_{\text{arena}}$ by computing entropy of the distribution $p(s^\tau|\bm{x}_{\text{arena}})$. A lower diversity indicates that when rats were in this location, they performed only a small variety of behaviors, and vice versa. Results for motifs at timescale $\tau=30$ are shown in Fig.\ref{fig:spatial heterogeneity}(b). The results indicate that when the rats were present near the center of the arena, their behavioral repertoire was more diverse. This was especially the case near ends of the restrainer (see Fig.\ref{fig:spatial heterogeneity} (a)), where the tracked rat could more directly interact with the trapped rat. Thus, the more diverse repertoire in the center may reflect the fact that more social behaviors were performed in the center than in the surround.

This led to the interesting question of whether some behavioral motifs were particularly preferred in certain locations. For instance, a social behavior may be preferred in the center, while a self-grooming behavior could be preferred in the surround. By looking at $p(s^\tau|\bm{x}_{\text{arena}})$, we found that indeed, some motifs were preferentially used near the center or the surround. Two examples with $\tau=30$ are shown in Fig.\ref{fig:spatial heterogeneity}(c). The motif on the left has a higher marginal probability near edges of the arena, and vice versa for the one on the right. This environmental dependence could be exploited to understand the ethological meanings of individual motifs.

\subsection*{Discovered behaviors and transition structure were ethologically meaningful}

%%%%%%%%%%%%%%%%%%%%%%%%%%%%%%%%%%%%%%%%%%%%%%%%%%%%%%%
%%%%%%%%%%%%%%%%%%%%%%%% FIGURE 7 %%%%%%%%%%%%%%%%%%%%%
%%%%%%%%%%%%%%%%%%%%%%%%%%%%%%%%%%%%%%%%%%%%%%%%%%%%%%%

In previous investigations using the restrainer paradigm (the experimental setup in the current study), prosocial behaviors between rats were studied \cite{bartal2011empathy,bartal2014pro,bartal2016anxiolytic,shan2016rodent, carvalheiro2019helping, tomek2019effects}. In short, if the free rat in the arena opens the restrainer door and releases the trapped rat repeatedly, this is determined to be a prosocial act. Further, in these investigations, a free rat is designated an “opener” if it opens from the trapped rat three times in a row among 12 testing sessions. We used the same paradigm and criterion to classify sessions used in the current study into those of openers ("prosocial rats", $N=63$) and those of non-openers ("non-prosocial rats", $N=105$). Importantly, behaviors analyzed here were recorded from a different paradigm where the restrainer could not be opened. Thus, door-opening was not present in our dataset and could not serve as a feature to distinguish prosocial and non-prosocial rats.

\begin{figure}%[tbhp]
\centering
\includegraphics[width=1\linewidth]{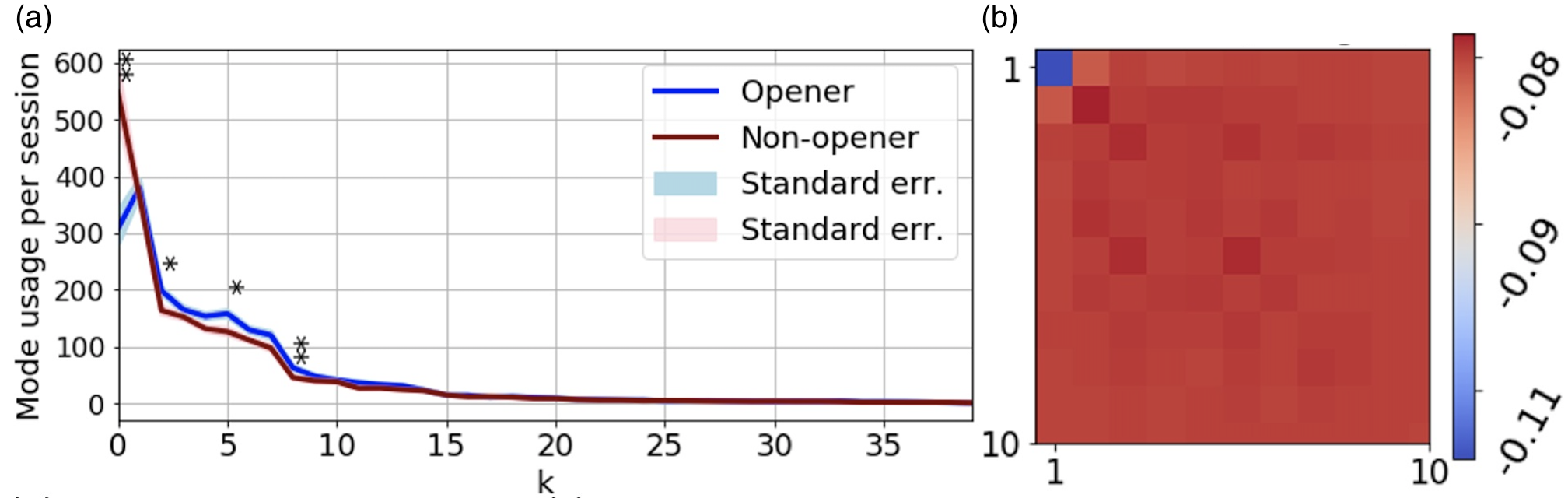}
\caption{Motif usage and transitions probabilities differed between prosocial rats and non-prosocial rats. (a) Marginal probabilities of motifs on $\tau=1$ timescale are significantlly different for some motifs. * shows $p<0.00125$; ** $p<0.00025$. Bonferroni corrected for multiple comparison. (b) Difference between transition matrices of prosocial and non-prosocial rats, also for $\tau=1$ motifs.}
\label{fig:opener nonopener}
\end{figure}

We asked if prosocial and non-prosocial rats showed different behavioral motif usage and transitions. We used Student's t test for significance and Bonferroni correction to counter multiple comparison error. Among 1-sec motifs, four are used significantly differently between the two groups (Fig.\ref{fig:opener nonopener} (a). Motifs are numbered by frequency. motifs \#1 and \#7, $p<0.01$; motifs \#6 and \#27, $p<0.05$; after Bonferroni correction). Similarly, we found that prosocial and non-prosocial rats have different transition probabilities (Fig.\ref{fig:opener nonopener}(b)). The spatial dependence of motif usage also differed between the two groups (data not shown). These results suggest that the motifs could serve as a behavioral signature of traits.

%%%%%%%%%%%%%%%%%%%%%%%%%%%%%%%%%%%%%%%%%%%%%%%%%%%%%%%%%
%%%%%%%%%%%%%%%% DISCUSSION %%%%%%%%%%%%%%%%%%%%%%%%%%%%%
%%%%%%%%%%%%%%%%%%%%%%%%%%%%%%%%%%%%%%%%%%%%%%%%%%%%%%%%%

\section*{Discussion}

Accurate description of behaviors is an important step towards linking behaviors to neural activities in the brain. Current models are mostly limited to a small timescale, in contrary to the multi-scale structures in real behaviors. Further, they did not account for how smaller motifs could compose larger, minute-scale motifs. As a result, while they are powerful tools for studying sub-second motor patterns, they are of little use for the majority of behaviors that live in supra-second domains. The current work aims to fill that gap. We identified four key challenges for motif identification and proposed an approach that addresses all.

We found that motifs in rat locomotion are indeed multi-scale. As our comparison between HMMs, GMMs and GMs showed, motifs with Markovian transitions could be discerned on scales ranging from 1 to 120 seconds. We found that the number of motifs decreased as timescale increased. This suggests that there may exist a sufficiently large timescale on which behaviors do not show motifs at all. Interestingly, we found that motif transition was closer to Markovian on smaller timescales. In other words, large-scale motif transition had a longer history dependence. We conjectured that this is because large-scale motifs reflect more goal-direction behaviors. This needs to be tested in further work. Unsurprisingly, we found that motif usage also depended on the environment. By carefully manipulating the environment and observing corresponding changes in motif usage and transitions, one may be able to link different stimuli to different motifs with our approach. Finally, we found that motif usage was a good feature for discriminating prosocial and non-prosocial rats. This raised the possibility that sensitive and non-invasive behavioral tests for internal states and traits could be developed with our approach.

There are several important limitations to our work. For simplicity, we assumed that motifs on each timescale are of the same duration. This was the most straightforward method to tokenize data. How this assumption can be relaxed is an important issue to address. Second, upon inspection, some motifs on each timescale appeared to be related to others -- they differ only by scaling or mirroring. They were classified as separate motifs by our models, even though they might represent different instances of the same motif. A model incorporating reflection invariance or scaling invariance may overcome this issue. Our preprocessing approach removed rotational and translational invariance; it could be extended to remove other invariance. Finally, our model does not explicitly assume interactions between motifs of different scales. One could design a directed acyclic graph (DAG) model that incorporates such interactions.

Our approach provides a general method to analyze multi-scale structures in animal behaviors. Such analysis is a good starting point for both experimental and computational investigations. It provides an unbiased, precise description of behavioral phenotypes. As a result, more minute responses to manipulations (e.g. genetic, pharmacological) can be discovered and studied. For computational works, it provides a descriptive model of behaviors that mechanistic models can be built upon. Biomechanical models that explain why rats have the repertoires that they do, and neural models of sequential generation of different motifs will be of particular interest.

\section*{Acknowledgements}

We would like to thank Nora Molasky for help with data collection. The authors declare no competing financial interests.

\section*{Broader impact statement}

We believe that animal researchers may benefit from our approach and insights. Our work does not directly or indirectly affect animal welfare in laboratories. At this moment we do not believe that our work has further societal consequences.

\end{document}